\documentclass[12pt,preprint]{aastex}

\shorttitle{X-ray and optical variability of NGC 1313 X-2}
\shortauthors{Mucciarelli et al.}

\begin{document}

\title{X-ray and optical variability of the ultraluminous X-ray source NGC 1313 X-2}

\author{P. Mucciarelli\altaffilmark{1,2}, L. Zampieri\altaffilmark{2}, 
A. Treves\altaffilmark{3}, R. Turolla\altaffilmark{4} and R. Falomo\altaffilmark{2}}
\altaffiltext{1}{Department of Astronomy, University of Padova, Vicolo 
dell'Osservatorio 2, I-35122 Padova,
Italy; paola.mucciarelli@oapd.inaf.it}
\altaffiltext{2}{INAF-Astronomical Observatory of Padova, Vicolo dell'Osservatorio 5,
I-35122 Padova, Italy; zampieri@pd.astro.it, falomo@pd.astro.it}
\altaffiltext{3}{Department of Physics and Mathematics, University of Insubria,
Via Valleggio 11, I-22100 Como, Italy; treves@mib.infn.it}
\altaffiltext{4}{Department of Physics, University of Padova, Via Marzolo 8, I-35131
Padova, Italy; turolla@pd.infn.it}

\begin{abstract}

We present an analysis of recent {\it XMM-Newton} and {\it HST}
archive data of the ultraluminous X-ray source NGC 1313
X-2. Quasi-simultaneous observations taken with {\it XMM-Newton}, {\it HST}
and VLT allow us to study both the X-ray light curve and its
correlation with the optical emission of the two proposed ULX
counterparts. At the end of December 2003 the source experienced a
short, but intense flare, reaching a maximum luminosity of
$\sim\,10^{40}$ erg/s. At the same time, the optical flux of both the suggested  
counterparts did not show pronounced variations ($\la 30$\%).
Assuming that the ULX emission is isotropic and taking X-ray
reprocessing into account, the optical data for one of the proposed
counterparts are consistent with it being an early type, main sequence
star of $\sim 10-18 M_\odot$ losing matter through Roche-lobe
overflow onto a $\sim 120 M_\odot$ black hole at an orbital separation
corresponding to a period of $\sim 2$ days.

\end{abstract}


\keywords{galaxies: individual (\objectname{NGC 1313}) --- stars: individual 
(\objectname{NGC 1313 X-2}) --- X-rays: binaries --- X-rays: galaxies}

\section{Introduction}

When, at the beginning of the 1980s, point-like, off-nuclear X-ray
sources in the field of nearby galaxies were first detected (see,
e.g., \citealt{fab89}), it was immediately recognized that their
luminosity was unusually large. If physically associated with their
host galaxies, these sources would have an isotropic luminosity
in excess of the Eddington limit for a $10 M_\odot$ object. Nowadays,
more than 150  ultraluminous X-ray sources (ULXs) are known (see
e.g. \citealt{rob00,col02,swa04,liu05}).

It is estimated that a significant fraction of ULXs are interacting
supernovae or background AGNs ($\sim$ 50\%; see
\citealt{fosc02,mas03,swa04}). However, the X-ray variability 
of many of them is similar to that observed in Galatic X-ray
binaries (see e.g. \citealt{lapar01,col02,swa04,zamp04}, hereafter Z04). 
The recent detection of a 
62 days modulation in the light curve of M 82 X-1, interpreted as the 
orbital period of the system, provided a direct confirmation of the 
binary nature of at least some ULXs \citep{ksm06a,ksm06b}. Moreover, ULX
spectral properties share similarities with those of Galactic black hole X-ray 
binaries (BHXRBs, e.g. \citealt{fosc02a}). In several cases the spectrum can be well
reproduced by a multicolor disk (MCD) blackbody plus a power law (PL),
although the temperature of the MCD component is often much lower than
that observed in BHXRBs (e.g. \citealt{mi03,mi04,fk05}). For the
brightest ULXs, a possible curvature above 2-3 keV has been recently
reported and more sophisticated spectral models appear to give better
agreement with observations \citep{stob06}.

All these properties along, in some cases, with the detection of
stellar-like optical counterparts (\citealt{rob01,goad02,liu02,liu04,kaa04}; Z04;
\citealt{kaa05,muc05}, hereafter M05; \citealt{sor05}), strongly suggest that a 
sizeable fraction of ULXs are accreting X-ray binaries. The present debate is focussed on
understanding what type of binaries they are.  Many of the ULX
properties can be explained if they do not emit isotropically
\citep{kin01,kin02,kp03} or are dominated by emission from a relativistic jet
\citep{kfm02,gak02,kaa03}. Another possibility is that they are truly
emitting above the Eddington limit for $10 M_\odot$, either because
accretion proceeds through a slim disk \citep{ebi03,kaw03} or because
the compact object is an Intermediate Mass Black Hole (IMBH) with a
mass in excess of $100 M_{\odot}$ (e.g. \citealt{cm99,mi03,pat05,pat06}).
Despite the inherent difficulties due to the low counting statistics, X-ray timing 
analysis has been attempted in some ULXs and led to the detection of a quasi
periodic oscillation in the power density spectrum of M 82 X-1. This may
represent a powerful, independent method to measure the black hole mass 
\citep{sm03,ft04,dew06,muc06}.

Multiwavelength observations are an unvaluable tool to investigate the
nature of ULXs. Radio emission, when present, gives important clues about
the geometry, energetics and lifetime of ULXs \citep{kaa03,miln05}.
Optical observations are crucial to identify ULX counterparts and
to study the properties of putative ULX binary systems. Up to now only a 
very
small number of ULXs have been convincingly associated with stellar
objects of known spectral type (e.g. \citealt{liu02,liu04,kaa04};
M05).  All these ULXs are hosted in young stellar environments or
star-forming regions and their optical counterparts have properties
consistent with those of massive stars. Some ULXs are also associated
with extended optical emission nebulae \citep{pak02,pak05}.

NGC 1313 X-2 is a well studied ULX. The X-ray variability, high (isotropic)
luminosity and presence of a soft X-ray spectral component  make it a
prototypical object. Furthermore, the presence of an emission nebula
(\citealt{pak02,ramsey06}; Z04) and the detection of optical counterpart(s)
(Z04; M05) provide a considerable amount of information on the ULX environment,
available only for a very limited number of objects. Here we present a
systematic study of the X-ray and optical variability of NGC 1313 X-2 based on
archive data of the {\it XMM-Newton} satellite, and the ESO VLT and {\it HST}
telescopes. Observations are reported in \S~\ref{obs} and results in
\S~\ref{variability}, where also a model for the optical emission is presented.
Discussion follows.

\section{Observations}\label{obs}

\subsection{X-ray observations}\label{x-ray}

{\it XMM-Newton} observed NGC 1313 in 9 exposures taken between 2000
and 2004. All the observations are listed in Table \ref{tab:olxmm}. A
detailed analysis of the October 2000 data of NGC 1313 X-2 have been
performed by Z04 (see also \citealt{tur04}). Here we report results from an
analysis of the EPIC pn exposures of the 8 more recent
observations. Data reduction and extraction have been carried out with
standard software ({\sc XMM-SAS} v. 6.0.0). All the observations are
affected by solar flares. The good time intervals left after
subtraction of the high background periods (when the total off-source
count rate above 10 keV is $>$ 1.0 counts s$^{-1}$ for EPIC pn) are
reported in Table
\ref{tab:olxmm}.
For the analysis we consider all the exposures with a good time
interval longer than 1 ks. 
After performing standard cleaning of the
event lists, we extracted source counts from a circle of 40$\arcsec$
centered on the position of NGC 1313 X-2 (Z04). The background
counts were extracted from a circle of 50$\arcsec$ on the same CCD.

The spectral analysis was carried out within {\sc XSPEC} (v.
11.2.0). A two-component model consisting of an absorbed 
multicolor disk blackbody (MCD) plus a power law (PL) has been employed
throughout, as it is routinely done for BHXRBs. Previous applications of
the same spectral model to a number
of ULXs, including NGC 1313 X-2 (\citealt{mi03,mi04,cr04,kong05}; Z04) gave
a satisfactory fit to the data. The absorbing column
density inferred from the different datasets is consistent with a
constant value. We then performed again the fits fixing $N_{H}$ 
to the average value weighted by the exposure time 
($N_{H} = 4.02 \times 10^{21}$cm$^{-2}$). For consistency, we also repeated the 
analysis of the 2000 EPIC pn spectrum of NGC 
1313 X-2 following the procedure described above. The results from the 
spectral fits are
reported in Table \ref{tab:x2fit}. The {\it XMM-Newton} fluxes were
consistently derived from the parameters of the spectral fits and are
reported, with the corresponding luminosities, in Table \ref{tab:x2fit}
(a distance of 3.7 Mpc was assumed for the host galaxy;
\citealt{tul88}). The errors have been estimated from the maximum and
minimum values of the flux, obtained varying the fit
parameters systematically.

\subsection{Optical observations}
\label{optical}

\begin{figure} 
\plottwo{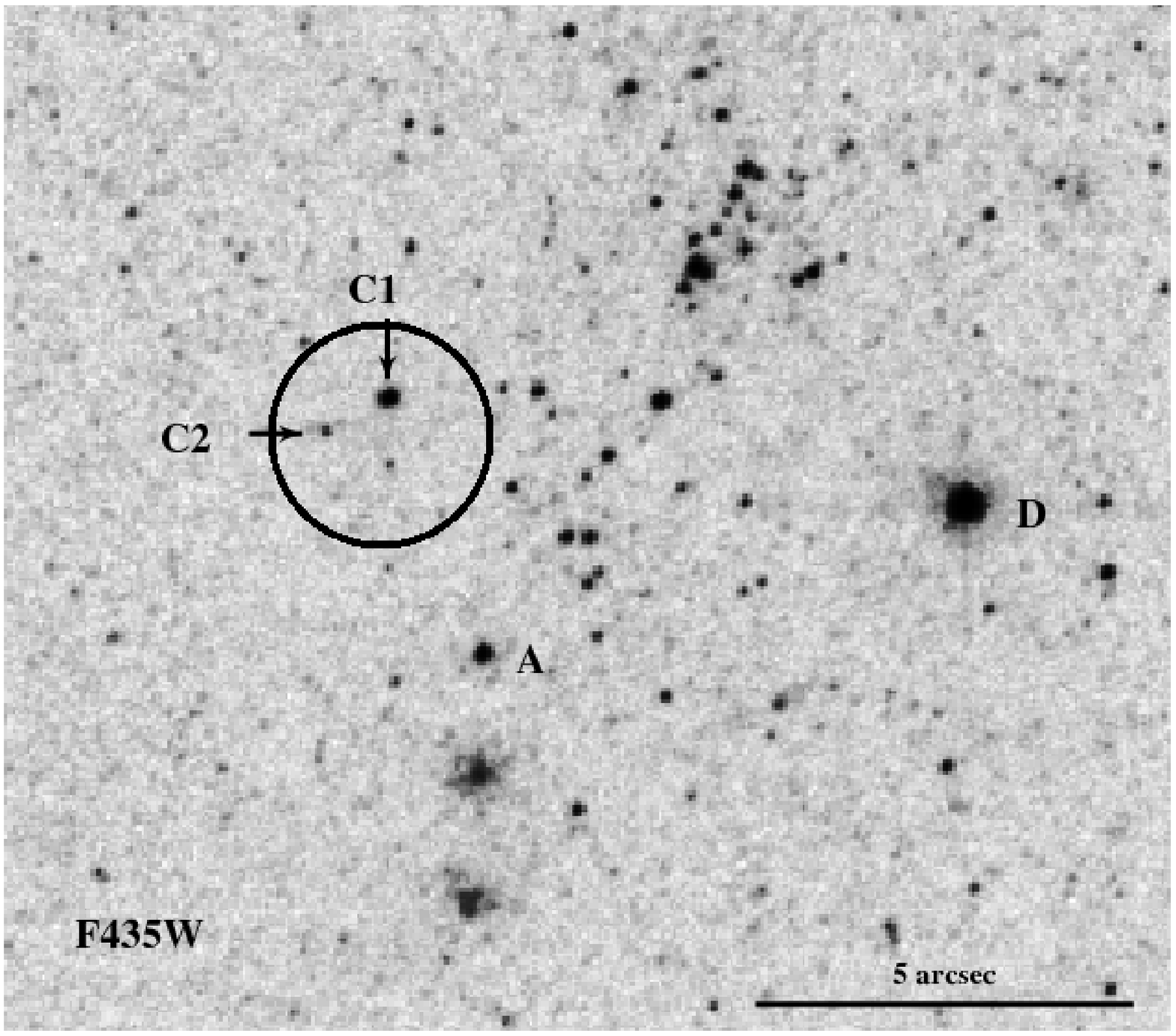}{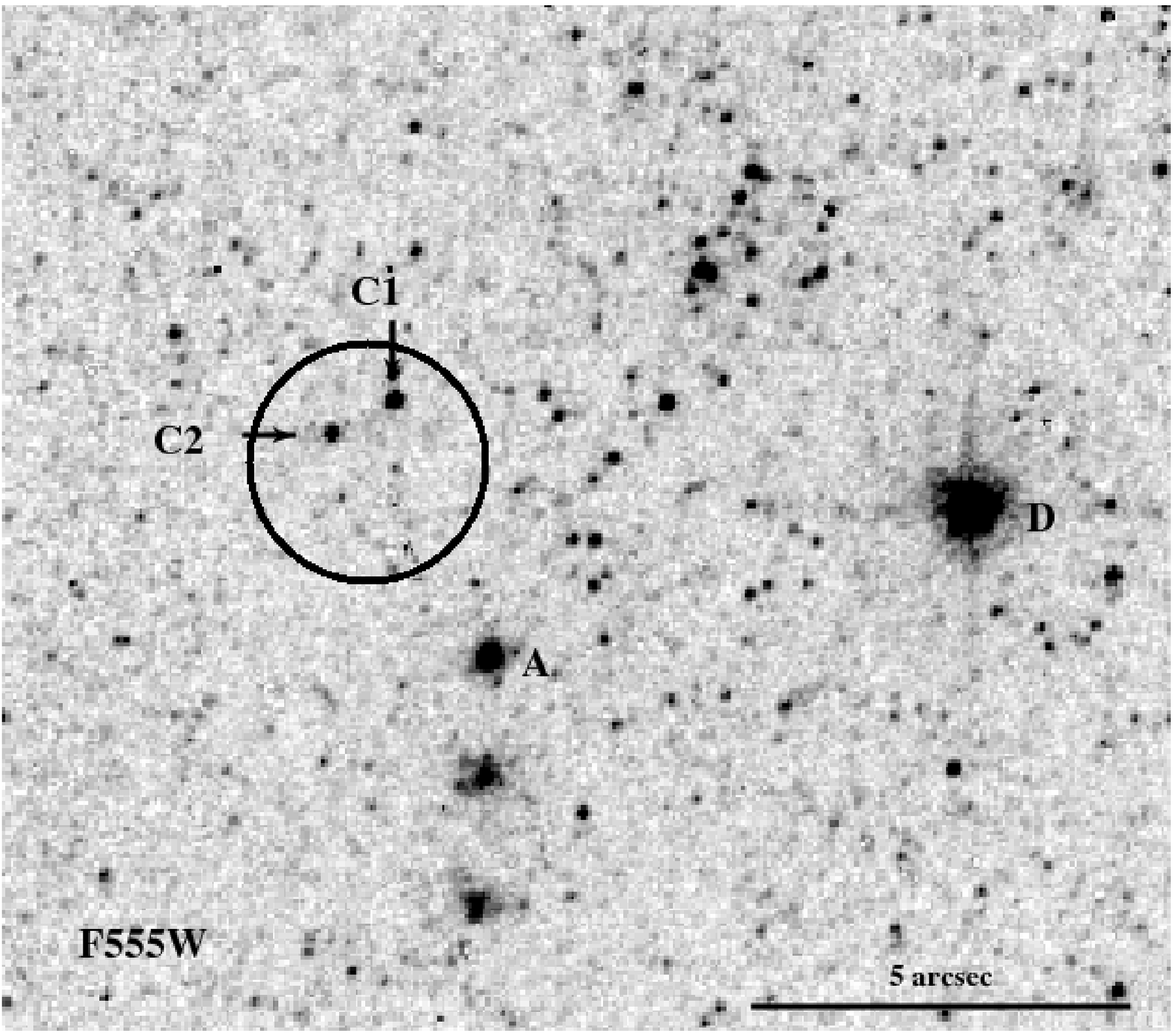} 
\caption{{\it HST}+ACS images of NGC 1313 X-2: left F435W (B) band, right 
F555W (V) band. The {\it Chandra} error box, the candidate optical counterparts C1 and C2 and the field sources A
and D are shown (following Z04 and M05).
\label{fig:bv}} 
\end{figure}

B, V and R images and spectra of NGC 1313 X-2 were taken with
VLT+FORS1 in December 2003. The results were presented by M05. {\it
HST} images of this field were also obtained with ACS in two epochs
(see Figure \ref{fig:bv}). The optical observations were performed in
parallel with the {\it XMM-Newton} pointings.  The observation log of the VLT
and {\it HST} images considered here are reported in Table
\ref{tab:olxmm}. Aperture photometry was performed on the drizzled
calibrated data (reduced by the {\it HST} pipeline) and transformed to
the Cousins system following \cite{sir05} (see Table
\ref{tab:lmag}).

The uncertainty on the {\it HST} magnitudes are dominated by the calibration
error ($\sim$ 0.03 mag for the {\it HST} photometry of point sources),
including filter transformation.
As a further check of the internal consistency of the {\it HST}
photometry, we compared the magnitudes of thirteen field stars
obtained in the two epochs. The difference is significant only for one
source in the sample ($\sim$ 0.3 mag). Excluding this source, the
variability of which is probably intrinsic, the magnitude changes are
randomly scattered around zero, with a mean absolute deviation of 0.04
mag.

The {\it HST} images clearly confirm that two distinct objects are
present inside the X-ray error box of NGC 1313 X-2 (see Figure
\ref{fig:bv}), as first shown by M05 on the basis of VLT data. 
If we assume $A_V \simeq 0.3$ (Z04), taking Galactic absorption into
account (the \citealt{card89} extinction law with $R_V=A_V/E(B-V)=3.1$
has been adopted throughout), the unreddened colors inferred from the
1st {\it HST} epoch are (B-V)$_0 \sim -0.13$ and $\sim$1.46 for C1 and
C2, respectively (see Table \ref{tab:lmag}). The color for object C1
is consistent with that derived from VLT data ((B-V)$_0 \sim -0.2$;
M05). On the basis of the {\it HST} photometry, the color of object C2
is close to that of a K3-K4 supergiant. For both objects, there is
evidence of variability in the V band between the two
{\it HST} epochs ($\sim 0.1$ mag, Figure \ref{fig:x2cl}; see also \citealt{ramsey06}).


\section{X-ray and optical variability}\label{variability}

\subsection{X-ray light curve}\label{lightcurve}

\begin{figure}
\begin{center}
\includegraphics[angle=-90,scale=.50]{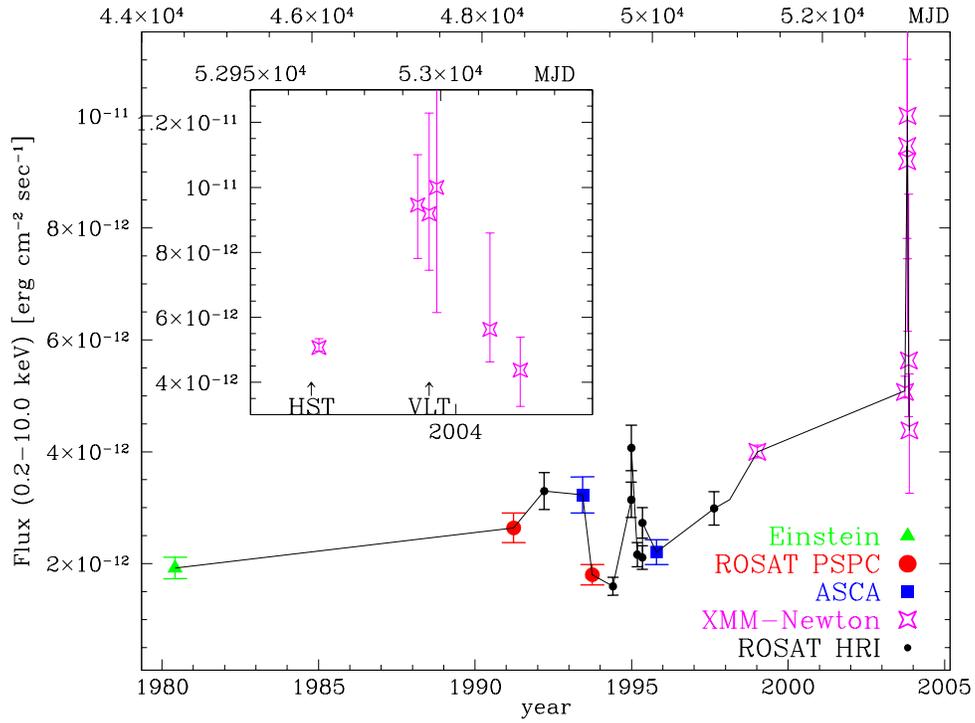}
\end{center}
\caption{The X-ray lightcurve of NGC 1313 X-2. Fluxes are unabsorbed and refer to the 
[0.2-10.0] keV energy interval (see Table \ref{tab:x2fit}). {\it
Einstein}, {\it ROSAT} and {\it ASCA} points are taken from Z04. The insert refers
to the more recent {\it XMM-Newton} data (Table \ref{tab:x2fit})
and the arrows mark the time of the {\it HST} and VLT observations (Table \ref{tab:olxmm}).
\label{fig:x2fl}}
\end{figure}

Figure \ref{fig:x2fl} shows the (unabsorbed) X-ray flux
for all the available observations of NGC 1313 X-2. The {\it XMM-Newton} data
were derived from the best fitting spectral models (see Table
\ref{tab:x2fit}), while the {\it Einstein},
{\it ROSAT} and {\it ASCA} data are taken from Z04.

Until 2000 NGC 1313 X-2 exhibited variability up to a factor of two on a
timescale of months, with a maximum luminosity of $\sim 4 \times
10^{39}$ erg s$^{-1}$ (Z04). 
Around December 25, 2003 (observation 6 in Table
\ref{tab:olxmm}), the source produced an intense
flare, reaching a maximum unabsorbed flux of $\sim 10.0 \times
10^{-12}$erg cm$^{-2}$ s$^{-1}$ (Table \ref{tab:x2fit}). At the
distance of NGC 1313 this corresponds to an intrinsic luminosity of
$\sim 10^{40}$erg s$^{-1}$. Clearly this value depends on the adopted
spectral model and hence should be taken with care (see e.g. the
slighty smaller values recently obtained for the October 2000 observation
by \citealt{stob06} adopting a more sophisticated
spectral model). We also measured the fluxes of another ULX in the
field (NGC 1313 X-3), known to be an interacting supernova (SN 1978K),
in order to check if the significant luminosity increase was real
or artificially produced by residual systematic effects between the
two observations. The flux of the supernova ($\sim 8.2 \times
10^{-13}$erg cm$^{-2}$ s$^{-1}$) is consistent
with a constant, within the uncertainties (the variation is $\la$
20\%). Hence we conclude that the luminosity increase of NGC 1313 X-2
is significant and fully qualifies NGC 1313 X-2 as a bright
ULX.

From the observed maximum luminosity ($L_{max} \sim 1.5 \times
10^{40}$erg s$^{-1}$) and assuming isotropic emission, the black hole
mass $M_{BH}$ obtained setting $L_{max}=L_{Edd}$ ($L_{Edd}$ is Eddington
luminosity) is $\simeq 120 M_\odot$, about 2 times larger than that
previously estimated by Z04. Sub-Eddington accretion would imply an
even larger mass.

\subsection{X-ray spectral changes}\label{spectchange}

The spectra of numerous ULXs, including NGC 1313 X-2, are well described by a
MCD+PL model, similarly to those observed in Galactic BHXRBs. In fact, there is
also some evidence of a closer similarity, inasmuch as some ULXs appear to show
state transitions \citep{mak04,win05}. In the following we summarize the main
results of an analysis of the X-ray spectral variability of NGC 1313 X-2. The
most significant result is that the slope of the PL component seems to
correlate with the flux, i.e. at higher fluxes the spectrum hardens (see Table
\ref{tab:x2fit}). This behavior was already noticed by Z04 on the basis of a
comparison between two {\it ASCA} observations and is opposite to that
usually shown by Galactic BHXRBs. A similar correlation was also
observed in a few ULXs in the Antennae galaxy by \cite{fab03}. The MCD
component is important in the Oct 2000 and in the 2003 pointings with
higher counting statistics (observations 4 and 5 in Table
\ref{tab:olxmm}). Although there is some evidence of intrinsic
variability of the thermal component, no definite conclusion can be
reached at present because of the insufficient statistics. Finally, we note
that the flux of the MCD component is comparable to that of the PL
component (see Table \ref{tab:x2fit}).

\subsection{Modeling the optical emission}\label{model}

In order to study the optical emission properties of objects C1 and C2 and
compare them with the {\it HST}+VLT photometry, we implemented a model to
compute the optical spectrum of a binary system with an IMBH taking irradiation
effects into account. Our calculation relies on the same assumptions discussed
in \cite{cop06}, who recently presented a thorough investigation of the
infrared-through-optical emission properties of X-ray binaries with IMBHs. More
specifically, we assume that accretion onto the IMBH is fuelled by a massive
companion filling its Roche lobe and that the X-ray emission is isotropic; the
consequences of introducing some degree of beaming are discussed later on. A
standard Shakura-Sunyaev disk (e.g.  \citealt{fkr02}) is assumed and both the
X-ray irradiation of the companion (including the effects of disk shadowing)
and the self-irradiation of the disk are accounted for.  Following
\cite{cop06}, radiative transfer at the donor and disk surfaces is treated
assuming a plane-parallel atmosphere in radiative equilibrium, illuminated by
the X-ray flux emitted from the innermost part of the accretion disk (see also
\citealt{wu01}). In order to keep our treatment simple, the companion
star is taken to be spherical, neglecting the effects produced by the Roche
lobe geometry and also those related to the (possible) deformation induced by
radiation pressure. Limb and gravity darkening were not included.

The model depends on the masses of the two components, the binary
period (which, in turn, fixes the orbital separation), the accretion
rate and the (unirradiated) temperature of the donor, in addition to
the inclination angle and the orbital phase. The accretion efficiency
and the albedo of the surface layers were chosen to be
$\eta=GM_{BH}/c^2 r_{in}=r_S/2 r_{in}=0.17$ (where $r_S$ is the
Schwarzschild radius and $r_{in}=3 r_S$ is the inner disk radius) and
$\alpha=0.9$ (e.g. \citealt{jpa96}). Following
\cite{cop06}, we took the hardness ratio $\xi=F_X(<1.5\, {\rm keV})/F_X(>1.5\, 
{\rm keV})=0.1$. The absorption parameters in the same two spectral
bands were selected as $k_s=2.5$ and $k_h=0.01$.  The V and B
magnitudes of the (irradiated) disk plus donor have been computed for
several values of the parameters of the binary.  Each sequence of
models, at fixed inclination angle $i$, accretion rate $\dot M$ and
donor mass $M$, corresponds to a track in the color-magnitude diagram
(CMD; B-V vs. V here) along which the BH mass varies. Only inclination
angles smaller than $\sim 70 ^{\circ}$ are considered because eclipsing
effects of the accretion disk on the donor (and viceversa) have not been
taken into account. On the other hand, from the available data there is no positive
evidence for eclipses of the X-ray source or the donor in NGC 1313 X-2.
The mass and luminosity class of the donor fix its (unirradiated) surface
temperature $T_{eff}$.
Different tracks have been obtained varying the orbital period $P_{orb}$
which, in turn, determines the Roche lobe radius. The maximum allowed
period is that for which the Roche lobe radius is equal to the donor radius.
The computed tracks are compared with the optical (unreddened) magnitudes
and colors of objects C1 and C2 for both the VLT and {\it HST} observations in
\S~\ref{application}.

\subsection{Objects C1 and C2}\label{application}

\begin{figure}
\epsscale{.70}
\plotone{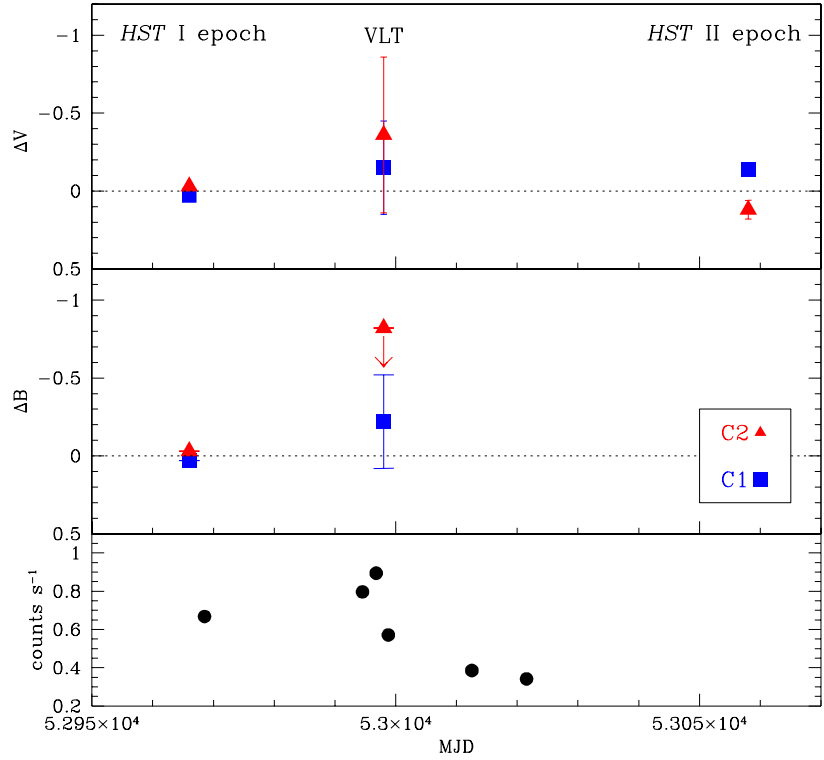}
\caption{{\it Top and middle panels}: $\Delta V=V-V(HST\; \rm{I\; epoch})$ and $\Delta 
B=B-B(HST\; \rm{I \;epoch})$
for objects C1 and C2 for the available epochs. The error bars 
correspond to 0.3 and 0.5 mag for V and B respectively (see text for 
details). {\it Lower
panel}: {\it XMM-Newton} count rates of NGC 1313 X-2 in the
[0.2-10.0] keV range. \label{fig:x2cl}}
\end{figure}

In order to constrain the parameters of the binary, we used the
optical binary emission model introduced in Section \ref{model}.  One
of the 2003 {\it XMM-Newton} pointings of NGC 1313 X-2 is within 2-3
days from the 1st {\it HST} epoch, while another is
close to the VLT observation. These two {\it XMM-Newton}
observations are those of 2003 November 25
and December 23; the latter was preferred to the observation of December 25 because
of the higher statistics. It is therefore of interest to compare the V
and B magnitudes of objects C1 and C2 in these two epochs. The variation of the unabsorbed
X-ray flux between the same epochs is $\sim$80\% (see Table \ref{tab:x2fit}).  At the same
time, however, the V and B band magnitudes of C1 do not show significant evidence of
variability. The relative photometric error between the {\it HST} and VLT data as measured
on a sample of field stars is in fact $\simeq 0.3$ mag (see Figure \ref{fig:x2cl}). A
similar  conclusion is reached also for object C2. Again, the magnitude change  between
the {\it HST} and VLT epochs is always smaller than the relative photometric error
($\simeq 0.5$ mag for objects fainter than V=24; see again Figure \ref{fig:x2cl}).

\begin{figure}
\epsscale{.8}
\plotone{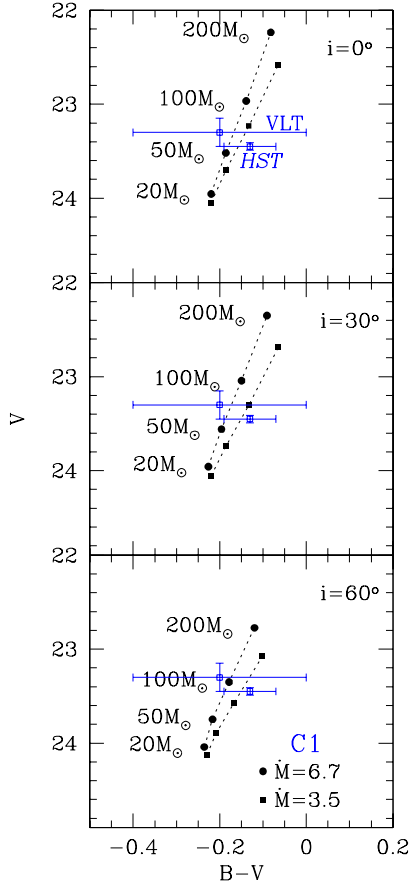}
\caption{Color-magnitude diagram for the (irradiated) disk plus donor model for
$P_{orb}\simeq 1.7$~d, $M\simeq 15 M_\odot$ and $T_{eff}\simeq
25000$~K (object C1). Each panel refers to a different inclination
angle $i$. The two tracks correspond to $\dot M=3.5$ and $6.7\, \dot
M_{Edd}$. These values are chosen in such a way to match the {\it XMM-Newton} flux
measured in the two observations of 2003 November 25 and December
23, that are quasi-simultaneous to the {\it HST} I
epoch and VLT observations, respectively.  The labels indicate the BH mass. 
The V magnitude and
B-V color as obtained from the VLT and the {\it HST} observations are also
shown ({\it open squares}).\label{fig:model}}
\end{figure}

As an initial guess for the donor parameters in our model we use the values inferred on
the basis of VLT photometry (M05):   an O9-B0 V star of $\sim 20 M_\odot$, $T_{eff}\sim
30000$ K for C1, and a G-K I star of $\sim 10 M_\odot$,  $T_{eff}\sim 4500$ K for C2.  As
mentioned above, this is consistent with what inferred from the {\it HST} photometry.
The donor star  in the case of of object C1 is assumed to be on the zero age main sequence
and in contact with the Roche lobe (i.e. its radius is equal to the Roche lobe radius).
Possible evolutionary effects or disturbances caused by the intense mass transfer are not 
accounted for. 
Results for object C1 are shown in Figure~\ref{fig:model} for two different values of $\dot
M$, chosen in such a way to match the 
{\it XMM-Newton} flux measured in the two observations of
2003 November 25 and December 23. The  tracks on the CMD diagram are in agreement with
the observed V band magnitude and (B-V) color of object C1 for 
$P_{orb}\simeq 1.7$~d, $M\simeq 15 M_\odot$ and $T_{eff}\simeq 25000$~K
(corresponding to an early B main sequence star). Taking into account for 
current uncertainties on both color and magnitude, the companion mass and 
temperature, and the orbital period may vary in the ranges $10\lesssim M/M_\odot\lesssim 18$,
$20000\, {\rm K}\lesssim T_{eff} \lesssim 30000\, {\rm K}$, and $1.5\lesssim P_{orb}
\lesssim 2$ d, respectively. 
If the donor makes contact with its Roche lobe and $0.1\lesssim\,M/M_{BH}\, \lesssim 0.8$, 
the orbital period 
becomes a function only of the donor radius (or mass). In fact, when combining
together the III Kepler's law and the expression for the Roche lobe
radius \citep{fkr02}, the dependence on $M_{BH}$ disappears.
As shown in Figure~\ref{fig:model}, the VLT
and {\it HST} photometric points intersect the corresponding tracks at
about the same value of the IMBH mass. This value increases with increasing inclination angle.
Results shown in Figure~\ref{fig:model} refer to orbital phase zero (superior conjunction). 
The variation in the V (B) band between 
the 1st {\it HST} epoch and the VLT one is $\simeq 0.23$ ($\simeq 0.25$), 
consistent (within the errors) with what observed. Thus, although
in these systems X-ray irradiation is very intense, the induced optical variability 
is not very large owing to the high intrinsic emission of the massive B donor.
The calculation for phases 0.25 and 0.5 gives results similar
to those obtained for phase 0, typically within 0.15 mag. Thus, this is the
expected amplitude of the modulation possibly induced by the orbital motion.
It is interesting to note that this result is consistent with the degree
of variability observed in the V band between the two {\it HST} observations
($\sim 0.1$ mag; see Figure \ref{fig:x2cl}).

A direct comparison of the three cases illustrated in  Figure~\ref{fig:model}
shows that relatively large values of the inclination angle ($i\gtrsim 
50^\circ-60^\circ$) are required in order to obtain the correct optical flux
for a BH mass $M_{BH}\sim 120 M_\odot$. At lower inclination angles (top and
middle panels of Figure~\ref{fig:model}),  the black hole masses needed to
reproduce the optical magnitudes and color of object C1 are too small for the
X-ray flux to be below the Eddington limit (if the emission is isotropic).
Therefore, unless the Eddington limit can be circumvented, the binary system is
expected to have a significant inclination angle. We note that, for any value
of $i$, the BH mass inferred from Figure~\ref{fig:model} is always $\gtrsim 70
M_{\odot}$. This absolute lower limits for $M_{BH}$ is still compatible  with a
very small beaming ($\gtrsim 0.6$). In this case the general  picture
discussed above reasonably continues to hold. On the other hand, if the ULX
emission is more seriously beamed, the BH mass could be smaller: for a beaming
factor $\sim$ 1/6, $M_{BH}$ can be as small as $20 M_\odot$ without exceeding
the Eddington limit. In this case, we expect  no X-ray irradiation of both the
disk and the companion, being the emission collimated away from the orbital
plane. To test this possibility we computed a new sequence of models, following
the same procedure outlined above, but switching off the disk/donor
irradiation. It turns out that it is possible to reproduce the correct
magnitude and color, although the donor is now less massive and cooler. 
However, this implies that the star is too small to fill its Roche lobe and
thus accretion  can not proceed through Roche lobe overflow. Wind accretion may
still be possible, although it seems unlikely that it can produce the required
value of $\dot M$.

The situation for object C2 is  reversed. We explored the
parameter space by varying the donor mass and orbital period, but did
not find any combination of values which could reproduce the data in the framework
of isotropic emission. In particular, X-ray irradiation causes the (B-V) color to always 
exceed the observed one. On the other hand, a massive and very extended K-type supergiant 
($M \sim 16 M_\odot$, $T_{eff}\sim 4000$~K, $P_{orb}\sim 800$ days) would have
properties consistent with those of object C2 if the black hole mass is $\sim 20
M_\odot$. This of course requires a (moderate) beaming. We checked that a beaming
factor of $\sim 1/6$ is enough and that the companion fills its Roche lobe. 
The optical magnitudes are correctly reproduced 
because the (unirradiated) disk contribution becomes negligible in comparison with 
the star intrinsic luminosity. However, in this case practically no variation in the optical
is expected in response to an increase of the accretion rate, and the predicted magnitudes
of C2 are constant. This is in contrast with the evidence of variability
observed in the V band between the two {\it HST} observations ($\sim 0.1$ mag; 
see Section \ref{optical} and Figure \ref{fig:x2cl}), although some variations may be induced also by the 
orbital ellipsoidal modulation of the donor (which we did not take into account). 


\section{Discussion}\label{concl}

Although present data do not allow to reach a definite conclusion on the actual
counterpart of the ULX NCG 1313 X-2, some firm points may be derived from 
the analysis presented in the preceeding sections. If C1 is the counterpart, as
it seems more likely, our model indicates that this is 
an IMBH X-ray binary with a relatively massive main sequence donor which 
fills its Roche lobe. Taking a
black hole mass of $\sim 120 M_\odot$, as required to account for the observed X-ray flux 
in terms of isotropic emission, the donor mass is in the interval $10-18 M_\odot$
(taking photometric uncertainties into account, \S~\ref{application}). This is
larger than the maximum main sequence mass of the parent stellar association, 
$\sim 8-9 M_\odot$, estimated using multicolor photometry and isochrone fitting by 
\cite{pak05} and \cite{ramsey06}. 
However, considering that the lower bound for the donor mass is $10
M_\odot$, the difference is small. We note also that, if C1 is the counterpart
and C2 belongs to the same stellar association, the estimated masses of the two
stars correctly places them on (or close to) and out of the main sequence,
respectively.
If the counterpart is C2 then the source is a binary formed by a late 
type, massive supergiant and a stellar mass black hole with beamed X-ray emission.
However, this scenario has some shortcomings. First, it predicts little if no optical 
variability, and this is in apparent contrast with the variations seen in 
the two {\it HST} observations. 
Second, the duration of the supergiant phase for a $\sim 15 M_\odot$ star is
very short (a few$\times 10^5 \, {\rm yr}$), making the possiblity of catching
the binary is such an evolutionary stage not very likely  \citep{pz06}. 
Third, the accurate astrometry of the field made possible by the cross-identification on the {\it
Chandra} and {\it HST} images of a background galaxy appears to strengthen the association of NGC 1313
X-2 with object C1, ruling out object C2 (J.-F. Liu, private communication).

For object C1, an orbital modulation of amplitude $\Delta V\sim 0.15$ is expected 
because of orbital inclination and X-ray irradiation effects. This modulation is superimposed
to a comparable variation caused by changes in the irradiating X-ray flux 
($\sim 0.2$ mag). In this respect, it is interesting to note that similar variations 
in the observed B band VLT+Subaru photometry of object C1 have been recently reported also 
by \cite{pak05}, consistent with our findings. In principle, with a sufficient and 
suitably spaced number of observations,
the orbital modulation can be singled out and measured with large area ground telescopes or
{\it HST}. The detection of this modulation would lead
to the unambiguous determination of the orbital period of the binary. This, in turn, 
would allow us to definitely discriminate between C1 and C2 and, most importantly, to
constrain the mass ratio of NGC 1313 X-2 and, eventually, the mass of the black hole.

\acknowledgements 
We are grateful to Claudio German\`a for his help with the binary X-ray
irradiation code. We also thank an anonymous referee  for useful comments that improved a
previous version of this paper. We acknowledge financial contribution from contract
ASI-INAF I/023/05/0 and MURST under grant PRIN-2004-023189. This paper is based on
observations obtained with {\it XMM-Newton}, an ESA science mission with instruments and
contributions directly funded by ESA Member States and NASA and on observations made with
the NASA/ESA Hubble Space Telescope, obtained from the data archive at the Space Telescope
Institute. STScI is operated by the association of Universities for Research in Astronomy,
Inc. under the NASA contract NAS 5-26555.

\clearpage

\begin{deluxetable}{llccrrc}
\tablecolumns{7}
\tabletypesize{\scriptsize}
\tablecaption{Observation log of the {\it XMM-Newton} EPIC pn pointings and of the
VLT+FORS1 and {\it HST}+ACS photometric observations of NGC 1313 
X-2.\label{tab:olxmm}}
\tablewidth{0pt}
\tablehead{ &\colhead{Instrument} & \colhead{Obs. Id.} & \colhead{Date} 
& 
\colhead{Exposure} & \colhead{GTI\tablenotemark{a}} & \colhead{Filter}
}
\startdata
1 & {\it XMM-Newton} EPIC pn  & 0106860101 & 2000-10-17 & 31637s & 20600s & Medium \\
2 & {\it XMM-Newton} EPIC pn   & 0150280101 & 2003-11-25 &  8365s & 1087s & Thin  
\\
3 & {\it XMM-Newton} EPIC pn  & 0150280201 & 2003-12-09 &  5620s &    0s & Thin  
\\
4 & {\it XMM-Newton} EPIC pn  & 0150280301 & 2003-12-21 & 10334s & 8272s & Thin  
\\
5 & {\it XMM-Newton} EPIC pn  & 0150280401 & 2003-12-23 & 14094s & 3600s & Thin  
\\
6 & {\it XMM-Newton} EPIC pn  & 0150280501 & 2003-12-25 & 15282s & 1668s & Thin  
\\
7 & {\it XMM-Newton} EPIC pn  & 0150280701 & 2003-12-27 & 16666s &    0s & Thin  
\\
8 & {\it XMM-Newton} EPIC pn  & 0150280601 & 2004-01-08 & 14756s & 7696s & Thin  
\\
9 & {\it XMM-Newton} EPIC pn  & 0150281101 & 2004-01-16 &  7034s & 4208s & Thin  
\\
& VLT+FORS1          &            & 2003-12-24 &840$\times$2 &  &B      
\\
& VLT+FORS1          &            & 2003-12-25 &600$\times$2 &  &V      
\\
& {\it HST}+ACS(I epoch) &         & 2003-11-22 &580$\times$2 & &F555w \\
& {\it HST}+ACS       &            & 2003-11-22 &630$\times$4 &  &F435w 
\\
& {\it HST}+ACS(II epoch) &        & 2004-02-22 &600$\times$4 &  
&F555w \\
\enddata
\tablenotetext{a}{Good Time Intervals in which the total off-source count rate above 10 keV is $<$ 1.0 counts s$^{-1}$.}
\end{deluxetable}

\begin{deluxetable}{ccccccccc}
\tablecolumns{9}
\tabletypesize{\scriptsize}
\tablecaption{Spectral analysis of {\it XMM-Newton} EPIC pn data of NGC 1313 
X-2.\tablenotemark{a}
\label{tab:x2fit}}
\tablewidth{0pt}
\tablehead{
& \colhead{Date}  & \colhead{Count rate} & \colhead{Flux\tablenotemark{b}} 
&
\colhead{$F_{MCD}/F_{PL}$\tablenotemark{c}} & 
\colhead{$L$\tablenotemark{d}} & \colhead{$kT_{MCD}$~[keV]} & 
\colhead{$\Gamma$} & \colhead{$\chi^{2}_{red}/d.o.f.$} 
}
\startdata
1 & 2000-10-17  & 0.24  &  4.00$^{+0.82}_{-0.62}$ &0.93 &  
6.52$^{+1.34}_{-1.02}$ & 
 0.16$^{+0.02}_{-0.01}$ & 2.3$^{+0.1}_{-0.1}$ & 1.24/75  
\\
2 & 2003-11-25  & 0.67  &  5.07$^{+0.28}_{-0.18}$ &--   &  
8.30$^{+0.04}_{-22.0}$
&  --                     & 2.3$^{+0.2}_{-0.2}$ & 1.06/44  
\\
4 & 2003-12-21   & 0.80  &  9.46$^{+1.55}_{-1.65}$ &0.63 & 
15.43$^{+2.53}_{-2.69}$ &
0.13$^{+0.01}_{-0.02}$ & 1.9$^{+0.1}_{-0.1}$ & 1.11/144 
\\
5 & 2003-12-23 & 0.89  &  9.19$^{+3.10}_{-1.74}$ &0.94 & 
14.99$^{+5.06}_{-2.84}$ 
&  0.15$^{+0.03}_{-0.03}$ & 1.8$^{+0.1}_{-0.1}$ & 0.99/140 
\\
6 & 2003-12-25   & 0.56  &  10.0$^{+8.80}_{-3.85}$ &0.87 &  
16.31$^{+14.36}_{-6.28}$ &
0.15$^{+0.04}_{-0.04}$ & 2.1$^{+0.2}_{-0.2}$ & 0.99/42  
\\
8 & 2004-01-08   & 0.39  &  5.63$^{+2.97}_{-1.00}$ &0.47 &  
9.18$^{+4.84}_{-1.63}$ &
0.13$^{+0.03}_{-0.02}$ & 2.5$^{+0.1}_{-0.1}$ & 0.82/67  
\\
9 & 2004-01-16   & 0.34  &  4.38$^{+1.01}_{-1.12}$ &0.45 &  
7.14$^{+1.65}_{-1.83}$ &
0.17$^{+0.02}_{-0.03}$ & 2.3$^{+0.1}_{-0.1}$ & 1.01/66  
\\
\enddata
\tablenotetext{a}{$N_{H}$ has been frozen at $4.02 \times 
10^{21}$cm$^{-2}$}
\tablenotetext{b}{Unabsorbed flux in units of $10^{-12}$ erg cm$^{-2}$ 
s$^{-1}$}
\tablenotetext{c}{Ratio of the (unabsorbed) fluxes of the MCD 
and PL components.}
\tablenotetext{d}{Unabsorbed luminosity, in units of $10^{39}$ erg s$^{-1}$, 
for a distance of 3.7 Mpc}
\end{deluxetable}

\begin{deluxetable}{lccc}
\tablecolumns{4}
\tabletypesize{\scriptsize}
\tablecaption{Observed magnitudes and colors of the two candidate optical 
counterparts of NGC 1313 X-2.\label{tab:lmag}}
\tablewidth{0pt}
\tablehead{
\colhead{ } &\colhead{Filter} &\colhead{C1}& \colhead{C2}
} 
\startdata
{\it HST} I epoch & B & 23.72$\pm$0.04 & 26.02$\pm$0.04 \\
VLT		  & B & 23.50$\pm$0.15 &  $\geq$ 25.2\\
{\it HST} I epoch & V & 23.75$\pm$0.04 & 24.46$\pm$0.04\\
VLT		  & V & 23.60$\pm$0.15 & 24.10$\pm$0.15\\
{\it HST} II epoch& V & 23.61$\pm$0.04 & 24.57$\pm$0.04\\
{\it HST} I epoch & B-V & -0.03$\pm$0.06 & 1.56$\pm$0.06 \\
VLT		  & B-V & -0.1$\pm$0.2 &  $\geq$ 1.1\\
\enddata
\end{deluxetable}

\clearpage

\end{document}